\documentstyle[preprint,pre,eqsecnum,aps]{revtex}
\begin{document}
\draft
\title{%
Jam phases in two-dimensional cellular automata model of traffic flow}
\author{Shin-ichi Tadaki\thanks{E-mail:tadaki@ai.is.saga-u.ac.jp}}
\address{Department of Information Science, Saga University,
Saga 840, Japan}
\author{Macoto Kikuchi\thanks{E-mail:kikuchi@godzilla.kek.jp}}
\address{Department of Physics, Osaka University, Toyonaka 560,
Japan}
\date{\today}
\maketitle
\begin{abstract}%
The jam phases in a two-dimensional cellular automata model
of traffic flow are investigated by computer simulations.
Two different types of the jam phases are found.  The
spatially diagonal long-range correlation obeys the power
law at the low-density jam configurations.  The diagonal
correlation exponentially decays at the high-density jam.
The exponent of the short-range correlation in the diagonal
direction is introduced to define the transition between
these two phases.  We also discuss the stability of the
jams.
\end{abstract}
\narrowtext
\section{Introduction}
The investigation of traffic flow has been based mainly on
the methods of fluid dynamics.  For example it has been
studied with the Burgers equation in one-dimensional cases.
Many attempts have been done to apply cellular automata
modeling to fluid for computational simplicity.  In recent
years, considerable interests have been developed also in
the cellular automata modeling of traffic flow.  One of the
simplest models of traffic flow in oneway expressways is the
rule-184 elementary cellular automaton\cite{wolfram}, which
is a simple asymmetric exclusion rule.  In spite of
simplicity of the model, it shows a phase transition from a
freely-moving phase at the low vehicle density to a jamming
phase at the high vehicle density.  More realistic models
accounting a variety of speed of cars and effects of
blockages have been investigated in one-dimensional
models\cite{nagel,nagel2,schadschneider,YKT}.
$1/f$ fluctuation has been observed in both
actual expressways\cite{musha} and
models\cite{nagel2,takayasu}.

Two-dimensional cellular automata models, however, have less
direct connection to real traffic flow problems.  It seems
to be an abstract model to a traffic system in a whole city
or an expressway network.  One of the simplest
two-dimensional traffic models has been investigated by
Biham, Middleton and Lovine (BML)\cite{biham}.  They found a
sharp transition between a freely-moving phase at the low
vehicle density and a jamming phases at the high density.
The model has been extended to take the probability of
changing directions of cars into account\cite{cuesta}.
Nagatani has studied the effect of a traffic accident or a
stagnant street to the growth of traffic
jams\cite{nagatani,nagatani2}.  He has also investigated the
model with two-level crossings\cite{nagatani4}.

These studies with cellular automata models have not been
concerning only the actual traffic problems or traffic light
controls; Attentions also have been paid to the phase
transitions and the self-organized behaviors on such
simple systems.  Most works have investigated the phase
transition of the system.  However, few have paid attention
to the structure and stability of the jam configurations.

In this paper we are interested mainly in the spatial
correlation in the jamming phase of the BML model, a
simple model for two-dimensional traffic flow, where cars are
represented as right and up arrows exclusively distributed
on a square lattice and controlled by a traffic light.  As
will be seen in later sections, there are two types of the
jam phases.  We discuss the spatial correlations of these
two phases and define the transition point. We also discuss
the stability of the jams in both phases.

This paper is organized as follows: We describe the BML
model in \S2, where the occurrence of the phase transition
from the freely-moving phase to the jam phase is mentioned as
well.  Spatial correlations in the jam phases are discussed
in \S3.  In \S4 we investigate the diagonal correlations.
The transition point between two type of jams is defined
using the short range correlation.  The stability of the jam
configurations is studied by applying a perturbation which
disturbs the jam configuration in \S5.  The distribution of
the lasting time of perturbations is discussed.  Section 6
is devoted to summary and discussion.

\section{The model}
The model we study is the same as the model-I of Biham,
Middleton and Levine\cite{biham}.  Cars are distributed on a
square lattice of $N\times N$ sites with periodic boundary
conditions both on horizontal and vertical directions.  Each
car is represented as an arrow directed up or right.  The
model is, therefore, a three state cellular automata with
empty sites, up-directed and right-directed cars as inner
states of each site.  A traffic light controls the dynamics,
such that the right arrows move only at odd time steps and
the up arrows move at even time steps (In the original BML
model, right arrows move at even steps and up ones at odd
steps.).  At odd time steps, each right arrow moves one site
to its right neighborhood if and only if the destination is
empty.  The corresponding rule is applied for up arrows.
The dynamics and the periodic boundary conditions guarantee
the conservation of the number of cars for each column and
row; There are $2N$ conservation rules.

The density of right (up) cars is given by
\begin{eqnarray}
p_{\rightarrow}&=&{n_{\rightarrow}\over N^2},\\
p_{\uparrow}&=&{n_{\uparrow}\over N^2},
\end{eqnarray}
where $n_{\rightarrow}$ ($n_{\uparrow}$) denotes the number
of right (up) arrows.
We examine the isotropic case where $p_{\rightarrow}=
p_{\uparrow} = p/2$ following Biham, Middleton and Levine.
They reported the existence of the
transition point $p=p_{\rm c}\sim 0.35$ (Nagatani has
investigated the transition point to be just under 0.4 in
the large system limit $N\rightarrow\infty$
\cite{nagatani3}.).  Below the transition point, all cars
move freely and the average velocity is $\overline{v}=1$.
All cars are blocked and the average velocity vanishes above
the transition point.  Such a sharp transition occurs
because the right and up arrows block each other.  That is
in contrast with one-dimensional models, where the average
velocity goes down to zero gradually with the increasing
density above the transition point.  In the following
sections we show results of the $N=128$ simulations.

\section{Two types of traffic jam}
Two types of traffic jam configurations are found in the BML
model through the simulation.  Figures \ref{ljam} and
\ref{hjam} show typical traffic jam configurations.
We investigate the structures, especially the spatial
correlations, of these two configurations.

In the low-density region, a single global cluster of jam is
oriented from the lower-left corner to the upper-right one.
The backbone of the jam lays diagonally and branches of the
jam spread horizontally and vertically like a herringbone.
It takes long time to reach the jam state starting from
random initial configurations. So the jam state at the low
density is far from random configurations and well
self-organized.  The situation seems to have similarity with
traffic jams caused by traffic accidents in the countryside.

In the high density region, on the other hand, patches of small local
clusters of jam cover the whole system. There are no
apparent global structure.  Starting from random initial
configurations, it needs only short time to get all cars stopped.
Thus randomness initially given is expected to remain.  This
type of configurations seems to model chronic traffic jams
in big cities.  An escape from a jam only means catching up
the tail of another jam in such a situation.

We investigate spatial correlation functions to study more
detailed characteristics of jam configurations.  We first
define the distribution function of right-directed
(up-directed) cars:
\begin{equation}
\rho_d(\vec{r}) = \sum_{i}\delta(\vec{r}-\vec{R}_{d,i}),
\end{equation}
where $d=\rightarrow\hbox{ or }\uparrow$ denotes the
directions and
\begin{equation}
\delta(x)=\cases{1&$x=0$,\cr 0&otherwise.}
\end{equation}
The coordinates $\vec{R}_{\rightarrow,i}$
($\vec{R}_{\uparrow,i}$ ) are those of the $i$-th right (up)
cars.  The correlation functions are defined as
\begin{equation}
C_{dd'}(\vec{r}) = {1\over n_d}
\left\langle\sum_{\vec{x}}\rho_d(\vec{x})\rho_{d'}(\vec{x}+\vec{r})
\right\rangle.
\end{equation}
The symbol $\langle\rangle$ denotes the sample average,
namely the average over the jam configurations started from
different random initial configurations.  The correlation
function $C_{dd'}(\vec{r})$, therefore, stands for the
probability to find a $d'$-directed arrow at the relative
position $\vec{r}$ from a $d$-directed arrow.  If the
correlation between the same directed cars vanishes,
$C_{dd}(\vec{r})$ goes down to the density $p_d$
($\vec{r}\not=0$).  We also define the {\it normalized}
correlation function
\begin{equation}
\tilde{C}_{dd}(\vec{r})={C_{dd}(\vec{r})-p_d \over 1- p_d}.
\end{equation}

The contour maps of the normalized correlation functions
$\tilde{C}_{dd}(\vec{r})$ are shown in Figs.~\ref{ljamc},
\ref{mjamc} and \ref{hjamc}.  At the low density (Fig.~\ref{ljamc}),
the correlation spreads diagonally over the system size.
The diagonally spreading spatial correlation corresponds to
the jam structure shown in Fig.~\ref{ljam}.  Increasing of
the density weakens the diagonal correlation.  The strong
diagonal correlation still remains at the intermediate
density (Fig.~\ref{mjamc}).  The decay of the diagonal
correlation causes the vibration in the anti-diagonal
direction.  At the high density (Fig.~\ref{hjamc}), at last,
the correlation is suppressed reflecting randomness.
\section{Defining the transition point}
The spatial correlation in the diagonal direction seems to
be a key to characterize the two types of jams.  The
diagonal correlation $\tilde{C}^{\rm diag}_{dd}(r)$ is
defined as the spatial correlation $\tilde{C}_{dd}(\vec{z})$
between diagonally separated arrows with the relative
position $\vec{z}=r(\hat{x}+\hat{y})$, where circumflex
identifies unit vector.  At the low-density jam, one can
find the diagonal correlation obeys the power law
(Fig.~\ref{correldiagl}).  At the high-density jam, the
long-range diagonal correlation dies out more rapidly than
power laws (Fig.~\ref{correldiagh}).

We investigate the exponents of the power law in the
diagonal correlation. It can be estimated easily for the
low-density jam as
\begin{equation}
\tilde{C}^{\rm diag}_{dd}(r)\sim r^{-\beta},\quad\beta\sim0.1.
\end{equation}
On the other hand, the correlation in the high-density jam
does not obey the power law.  So we approximate the
short-range correlation of the power law and define the
effective exponent for the high-density jam.  Actually we
fit the short-range data $(\log r, \log\tilde{C}_{dd}^{\rm
diag}(r))$, $(1\le r\le n< N/2)$ to a linear function by the
method of least squares.  Namely, parameters $\beta$ and
$\gamma$ are chosen to minimize a squared deviation
\begin{equation}
\chi_n(\beta,\gamma)^2=\sum_{r=1}^n
\left(\log\tilde{C}_{dd}^{\rm diag}(r) -\beta\log r -\gamma\right)^2,
\end{equation}
for fixed $n$. Then maximize the range $n$ under the condition
\begin{equation}
\left({1\over n}\min\chi_n(\beta,\gamma)^2\right)^{1/2}<\epsilon,
\label{fitcondition}
\end{equation}
where $\epsilon$ is an adequate constant, which all data in the
low-density phase are fit with.

Figure \ref{ap} shows the dependence of the exponent $\beta$
on the density $p$ for $N=64$, 128 and 200 systems.  One can
find a clear transition between low-density and high-density
regions at $p_t=p\sim0.52$.  The low-density jam above the
transition point $p_c$, namely $p_c<p<p_t$, shows the power-law
diagonal correlation.  The exponent of the correlation
is $\beta\sim0.1$.  At the high-density jam above the second
transition point, $p>p_t$, the diagonal correlation decays.
The exponent $\beta$ for the short-range correlation
increases in proportional to the density $p$ in the high-density
region.  The values of the exponent hardly depend on
the system size.  The values of the exponent themselves,
however, seem not to be meaningful for the high-density
region, because they depend on the value of $\epsilon$ in
Eq.~\ref{fitcondition}.

The diagonal correlation can be fit also with exponential
functions $\tilde{C}^{\rm diag}_{dd}(r)\propto\exp(-r/\xi)$
in for large $r$.  The correlation length $\xi$
exceeds the system size $N$ in the-low density jam region
(Fig.~\ref{xi}).  This causes the power-law dependence of
the diagonal correlation.  The correlation length suddenly
decreases to $\xi\ll N$ near the transition point $p_t$.
The estimated correlation lengths contain errors
due to statistical errors and finite-size effect
especially in the high-density jam phase.

\section{Stability of the jam}
Finally we investigate the difference of the stability of
the jam configurations in both phases mentioned above.  For
this end, we perturb the jam state to study the stability of
the phase after the jamming state appears.  In the center of
jams, there are blockade pairs with an up arrow blocking the
forehead of a right one or a right arrow blocking an up one.
To perturb the jam phase, we remove one of such blockade
pairs to the tailend of jams leaving a pair of vacant sites
in the center of the jam.  Then the jamming state melts
using this vacant sites.  All cars stop again after some
steps $t$ in the unit of a cycle of traffic light.  We
observe the distribution $P(t)$ defined as
\begin{equation}
P(t) = {n(t)\over\sum_{t=0}^\infty n(t)},
\end{equation}
where $n(t)$ is a number of events that the effect of the
perturbation remains until $t$.

Let us briefly describe the actual simulation on the model
to estimate the distribution $P(t)$.  For the given values
of $N$ and $p$, we first give an initial random state using an
adequate pseudo-random number generator.  Starting from
this state the system runs until a jam occurs.
Then the perturbation mentioned above is
applied, namely removing a randomly chosen blockade pair to
the tailend of the jam.  For example, we consider the case
that a blockade is a pair of a right arrow on $(i,j)$-site
and an up on $(i+1,j)$-site.  Then we start to find an empty
site from the left-nearest site to the right-next-nearest of
$(i,j)$-site leftward.  If $(i-k,j)$-site is empty, it is
marked as a candidate of the destination for the right arrow
on $(i,j)$-site to move.  The same procedure is applied also to
the up arrow on $(i+1,j)$-site, searching an empty site
downward and marking a candidate.  If a pair of vacant sites are
marked as the destination, each one of the blockade pair is
removed to the marked sites.  If there is no available pair
of empty sites, another blockade pair is tried to find for
removing.  After removing a blockade pair, the system runs
until the next jam occurs.  Then the next perturbation is given.
For one initial random state, $10,000$ such perturbations are
applied repeatedly. These $10,000$ data are obtained in one
run of the simulation.  The detailed method to remove a
blockade pair would not affect the results qualitatively.

The distributions $P(t)$ at the low-density jam are shown in
Fig.~\ref{plow}.  The effect of the perturbation remains for
long time.  It propagates along the sequences of blockade
pairs which compose the center of the jam.  So the
distribution $P(t)$ has a peak corresponding to the system
size (Fig.~\ref{plow}).

At the high-density jam (Fig.~\ref{phigh}), on the other
hand, the effect of the perturbation vanishes in shorter
times than the low-density cases.  The distribution $P(t)$
decays in the large $t$ region. The decay rate of $P(t)$
is not clear at this stage.

\section{discussion}
We studied the properties of the jam in a two-dimensional
cellular automata model of traffic flow.  The basic model
treated here is the same as the model-I of Biham, Middleton
and Levine\cite{biham}.  We found two types of the jam
(Fig.~\ref{phase}).  At the low density above the transition
($p_c<p<p_t$), the spatially diagonal long-range correlation
appears.  The backbone of the jam, sequences of blockade
pairs, lays diagonally through the whole system.  The
branches of the jam spread horizontally and vertically.
Thus the jam configurations are well-organized. We call this
type of jams as {\it self-organized jam}.  At the
high-density jam ($p>p_t$), on the other hand, local
clusters of jam cover the whole system and no global
structure remains.  We call this type of jams as {\it random
jam}.

The spatial correlations were studied to characterize these
two type of jam configurations.  The spatially diagonal
correlation obeys the power law at the low-density jam.
Spatial correlations decay with increment of the density.
We defined the short range exponent of the diagonal
correlation for the high-density jam.  The transition point
$p_t$ could be defined by the change in the values of the
exponent.

The long-range diagonal correlations decay exponentially
reflecting the randomness of the system.  The correlation
length $\xi$ exceeds the system size $N$ in the low-density
jam phase ($p<p_t$).  It is expected that the correlation
length remain finite in the $N\rightarrow\infty$ limit.  We,
however, have no clear evidence of the finite correlation
length in the limit.
Another possibility is that the correlation length diverges in the
thermodynamic limit.  In this case, it is expected that the two
transition points $p_c$ and $p_t$ coincide with each other, and the
low-density jam phase will be observed only just at $p_c$.

The stability of these jam configurations was investigated
by applying perturbations to the jam configurations.  We
investigated the distribution $P(t)$ of the time that the
effect of the perturbation remains.  The distribution $P(t)$
at the low density above the transition has a peak
corresponding to the system size reflecting the long-range
correlation. Detailed analysis of the distribution $P(t)$ is
discussed elsewhere.

The power-law relations of the diagonal correlation remind
us of the self-organized critical properties of the jam
configurations\cite{bak}.  The jam configurations seem to be
self-organized well at the low-density jam.  At the high
density, however, the initial randomness remains and thus
the jam configurations are not self-organized.  We have no
explanation on the origin of the power law and the values of
the exponent at present.

The occurrence of the jam may be affected by boundary
conditions and isotropy especially at the low density.  The
effect of anisotropy ($p_{\rightarrow}\not=p_{\uparrow}$)
has been studied by Nagatani\cite{nagatani3}.  He pointed
out that strong anisotropy prevents the system from the jam.
The effect of another isotropy is still unclear, namely the
effect of the equality of the horizontal and vertical system
sizes.

Cellular automata models of traffic flow are not restricted
to concern traffic flow problems.  They are simplified
abstract models of exclusion processes.  On one-dimensional
models analogies to the ballistic deposition have been
found\cite{biham,meakin,krug,janowsky}.  Although any
connection of the two-dimensional models to physical systems
have never been discussed so far, studies on these systems
are expected to clarify behaviors of complex systems.

\acknowledgements

The authors would like to appreciate T. Nagatani for sending
us his articles.

\begin{figure}
\caption{A typical jamming configuration in the low-density
region above the transition $p_c$. The global cluster of jam
is oriented diagonally.  The system size is $32\times32$ and
$p=408/(32\times32)\sim0.4$.}
\label{ljam}
\end{figure}
\begin{figure}
\caption{A typical jamming configuration in the high-density
region above the transition $p_c$. The system is covered by local
clusters of jam.  The system size is $32\times32$ and
$p=920/(32\times32)\sim0.9$.}
\label{hjam}
\end{figure}
\begin{figure}
\caption{The normalized correlation functions
$\tilde{C}_{dd}(\vec{r})$ at the low density.  The
upper diagram is of
$\tilde{C}_{\rightarrow\rightarrow}(\vec{r})$ and the lower
$\tilde{C}_{\uparrow\uparrow}(\vec{r})$.  The correlation
spreads diagonally over the system size.  The high value
regions at the upper-left and lower-right corners are caused
by the finite-size effects.  The system size is
$128\times128$ and $p=6552/(128\times128)\sim0.4$.  The average is
taken over 10 samples.}
\label{ljamc}
\end{figure}
\begin{figure}
\caption{The normalized correlation functions
$\tilde{C}_{dd}(\vec{r})$ at the intermediate density.  The
upper diagram is of
$\tilde{C}_{\rightarrow\rightarrow}(\vec{r})$ and the lower
$\tilde{C}_{\uparrow\uparrow}(\vec{r})$.  The correlation
mainly spreads diagonally over the system size.  There
appears the vibration in the anti-diagonal direction in
contrast to the low-density jam.  The system size is
$128\times128$ and $p=6552/(128\times128)\sim0.4$.  The average is
taken over 10 samples.}
\label{mjamc}
\end{figure}
\begin{figure}
\caption{The normalized correlation functions
$\tilde{C}_{dd}(\vec{r})$ at the high-density.  The
upper diagram is of
$\tilde{C}_{\rightarrow\rightarrow}(\vec{r})$ and the lower
$\tilde{C}_{\uparrow\uparrow}(\vec{r})$.  The values of
$\tilde{C}_{dd}(\vec{r})$ are almost zero over the whole
region except the vicinity of the coordinate origin.  The
system size is $128\times128$ and $p=14744/(128\times128)\sim0.9$.
The average is taken over 10 samples.}
\label{hjamc}
\end{figure}
\begin{figure}
\caption{The diagonal correlation
$\tilde{C}^{\rm diag}_{\rightarrow\rightarrow}(r)$, the
normalized correlation between right arrows separated
diagonally at the distance $2^{1/2}r$, are plotted for
$p\le0.6$.  The average is taken over 10 samples as the same
as Figs.~\protect{\ref{ljamc}} and \protect{\ref{hjamc}}.
The lines are fit with the method of least squares with
$\epsilon=0.05$ in Eq.~\protect{\ref{fitcondition}}.}
\label{correldiagl}
\end{figure}
\begin{figure}
\caption{The diagonal correlation
$\tilde{C}^{\rm diag}_{\rightarrow\rightarrow}(r)$, the
normalized correlation between right arrows separated
diagonally at the distance $2^{1/2}r$, are plotted for
$p\ge0.6$.  The average is taken over 10 samples as the same
as Figs.~\protect{\ref{ljamc}} and \protect{\ref{hjamc}}.
The lines are fit with the method of least squares for small
$r$ with $\epsilon=0.05$ in
Eq.~\protect{\ref{fitcondition}}.}
\label{correldiagh}
\end{figure}
\begin{figure}
\caption{Dependence of the exponents $\beta$ on the density $p$.}
\label{ap}
\end{figure}
\begin{figure}
\caption{Diagonal correlation length $\xi$.
It reaches the system size $N$ at $p=0.57$.}
\label{xi}
\end{figure}
\begin{figure}
\caption{
The distribution $P(t)$ at $p=6552/(128\times128)\sim0.4$ and
$N=128$.  The data corresponds to the summary of 3 runs, one
run means $10,000$ data as mentioned in the text.  It has a
peak corresponding to the system size.  }
\label{plow}
\end{figure}
\begin{figure}
\caption{
The distribution $P(t)$ at $p>0.6$ and $N=128$.  Each data
corresponds to the summary of 3 runs, one run means $10,000$
data as mentioned in the text.}
\label{phigh}
\end{figure}
\begin{figure}
\caption{Schematic
phase diagram of two-dimensional traffic flow model.}
\label{phase}
\end{figure}

\begin{references}
\bibitem{wolfram}S. Wolfram, Rev. Mod. Phys. {\bf55}, 601 (1983).
\bibitem{nagel}K. Nagel and M. Schreckenberg,
J. Phys. I France {\bf2}, 2221 (1992).
\bibitem{nagel2}K. Nagel and H. J. Herrmann,
{\it Deterministic models for traffic jam},
HLRZ preprint 46/93 (1993).
\bibitem{schadschneider}A. Schadschneider and M. Schreckenberg,
J. Phys. A{\bf 26}, L679 (1993).
\bibitem{YKT}S. Yukawa, M. Kikuchi and S. Tadaki,
{\it Dynamical phase transition in one dimensional traffic
flow model with blockage}, J. Phys. Soc. Jap. {\bf 63} (1994) in press.
\bibitem{musha}T. Musha and H. Higuchi,
Jap. J. Appl. Phys. {\bf15}, 1271 (1976).
\bibitem{takayasu}M. Takayasu and H. Takayasu, Fractals {\bf1},
860 (1993).
\bibitem{biham}O. Biham, A. A. Middleton and D. Levine,
Phys. Rev. A{\bf46}, 6124 (1992).
\bibitem{cuesta}J. A. Cuesta, F. C. Mart{\'\i}nez, J. M. Molera
and A. S\'anchez, Phys. Rev. E{\bf48}, 4175 (1993).
\bibitem{nagatani}T. Nagatani, J. Phys. Soc. Jap. {\bf62}, 1085 (1993).
\bibitem{nagatani2}T. Nagatani, Physica A{\bf198}, 108 (1993).
\bibitem{nagatani4}T. Nagatani, Phys. Rev. E{\bf48}, 3290 (1993).
\bibitem{nagatani3}T. Nagatani, J. Phys. Soc. Jap. {\bf62}, 2625 (1993).
\bibitem{bak}P. Bak, C. Tang and K. Wiesenfeld,
Phys. Rev. Lett. {\bf 59}, 381 (1987), Phys. Rev. A{\bf38}, 364 (1988).
\bibitem{meakin}P. Meakin, P. Ramanlal, L. M. Sander and R. C. Ball,
Phys. Rev. A{\bf34}, 5091 (1986).
\bibitem{krug}J. Krug and H. Spohn, Phys. Rev. A{\bf38} 4271 (1988).
\bibitem{janowsky}S. A. Janowsky and J. L. Lebowitz, Phys. Rev.
A{\bf45} 618 (1992).
\end{references}
\end{document}